\documentclass{eas}
\usepackage{graphicx}
\newcommand{\z}{$z$}

\newcommand{\HI}{H$\,${\small I}}

\newcommand{\spitzer}{{\em Spitzer}}
\newcommand{\micron}{$\mu$m}
\newcommand{\arcmin}{\prime}
\newcommand{\msun}{M$_\odot$}

\newcommand{\aap}{A\&A}
\begin{document}

\title{Extraplanar Dust in Spiral Galaxies: Tracing Outflows in the
  Disk-Halo Interface}
\runningtitle{Howk: Extraplanar Dust}
\author{J. Christopher Howk}
\address{Dept. of Physics, Univ. of Notre Dame, Notre Dame, IN 46556, USA}
\begin{abstract}

  There is now ample evidence that the interstellar thick disks of
  spiral galaxies are dusty.  Although the majority of extraplanar gas
  in the first few kiloparsecs above the plane of a spiral galaxy is
  matter that has been expelled from the thin disk, the
  feedback-driven expulsion does not destroy dust grains altogether
  (and there is not yet any good measure suggesting it changes the
  dust-to-gas mass ratio).  Direct optical imaging of a majority of
  edge-on spiral galaxies shows large numbers of dusty clouds
  populating the thick disk to heights $z\sim2$ kpc.  These
  observations are likely revealing a cold, dense phase of the thick
  disk interstellar medium.  New observations in the mid-infrared show
  emission from traditional grains and polycyclic aromatic
  hydrocarbons (PAHs) in the thick disks of spiral galaxies.  PAHs are
  found to have large scale heights and to arise both in the dense
  dusty clouds traced through direct optical imaging and in the
  diffuse ionized gas.  In this contribution, we briefly summarize
  these probes of dust in the thick disks of spiral galaxies.  We also
  argue that not only can dust can be used to trace extraplanar
  material that has come from within the thick disk, but that its
  absence can be a marker for newly accreted matter from the
  circumgalactic or intergalactic medium.  Thus, observations of dust
  can perhaps provide a quantitative measure of the importance of
  ``outflow versus infall'' in spiral galaxies.

\end{abstract}
\maketitle
\section{Introduction}


Extraplanar gas in spiral galaxies is a result of several important
phenomena.  The thick gaseous disks of galaxies (or even the more
extended halos or coronae) are in part produced by the outflow of
matter from the thin disk below, with the outflow energy provided by
stellar feedback.  There is, however, significant interest in the
possibility that some thick disk material may have come from outside
of the galaxy (see recent review by Sancisi et al. 2008).  The
continued formation of stars in galaxies (both individual galaxies and
galaxies on the whole) and the observed metallicity distribution of
long-lived stars in our own Galaxy almost require the continued infall
of low-metallicity matter from outside of the star forming disks of
spirals (Larson 1972, Chiappini et al. 1997, Wolfe et al. 2005),
although direct observational evidence for this is difficult to come
by.  The raw materials for such infall may come from primordial
intergalactic gas or near-primordial circumgalactic gas, such as high
velocity clouds (HVCs) condensed from a galactic corona (e.g,. Peek et
al. 2008).  While matter stripped from dwarf galaxies may contribute
to some extent (as evidenced by the Magellanic Stream), there are
arguments against dwarf galaxies as the primary source of material for
massive spiral galaxies (Peek 2009; Sancisi et al. 2008).

What observational evidence is there for such accreting matter,
especially accreting neutral matter that may be readily incorporated
into the star forming gas of the thin disk?  The contributions by
Oosterloo and Fraternali to these proceedings provide some direct and
indirect evidence for neutral material at anomalous velocities near
galaxies.  In particular, the counter-rotating clouds seen near the
edge-on spiral NGC 891 are compelling examples of this infall
(Oosterloo et al. 2007).  However, they also argue that the rotation
curves of extraplanar \HI\ provide indirect evidence for the accretion
of matter onto NGC~891 and other galaxies.

Here we suggest another method for studying the mixture of outflowing
and infalling gas in the interstellar thick disks of galaxies: using
the gas-to-dust ratio traced by the ratio of infrared emission to \HI\
21-cm column density (see \S \ref{sec:dust2gas}).  By outflowing in
this case, we mean material ejected from the thin disk, even if it
will eventually return, for example, as part of a galactic fountain.
Infalling refers to low-metallicity matter from the local
circumgalactic or intergalactic matter.  The implicit assumption is
that dust in the thick disk is exclusively linked to enriched material
expelled from the underlying thin disk.  Thus, dust becomes a tracer
particle that marks material that has come from within a galaxy, while
its absence is an indicator of (nearly) primordial material in the
thick disk.

\section{Thick Disk Dust in Spiral Galaxies}

There is no doubt that the energetic process that expel gas from the
thin disks of spiral galaxies do so in a manner that does not
altogether destroy dust grains.  There is strong evidence for
extraplanar dust in the Milky Way: depletion studies of warm halo gas
(Sembach \& Savage 1996) and warm ionized halo gas (Howk \& Savage
1999b, Howk et al. 2006), as well as infrared emission from the
high-\z\ neutral Galactic cirrus and potentially the warm ionized
medium (Lagache et al. 2000, Odegard et al. 2007) all provide evidence
for extraplanar dust in our Galaxy.  Dusty chimneys and ``worms''
stretching to high-\z\ have been studied for quite some time (Koo et
al. 1992), and the walls of large supershells show thermal dust
emission (e.g., Callaway et al. 2000).  

\subsection{Cold Clouds Traced by Dust in the Thick Disks of Spirals }

\begin{figure}[!ht]
\includegraphics[width=12cm]{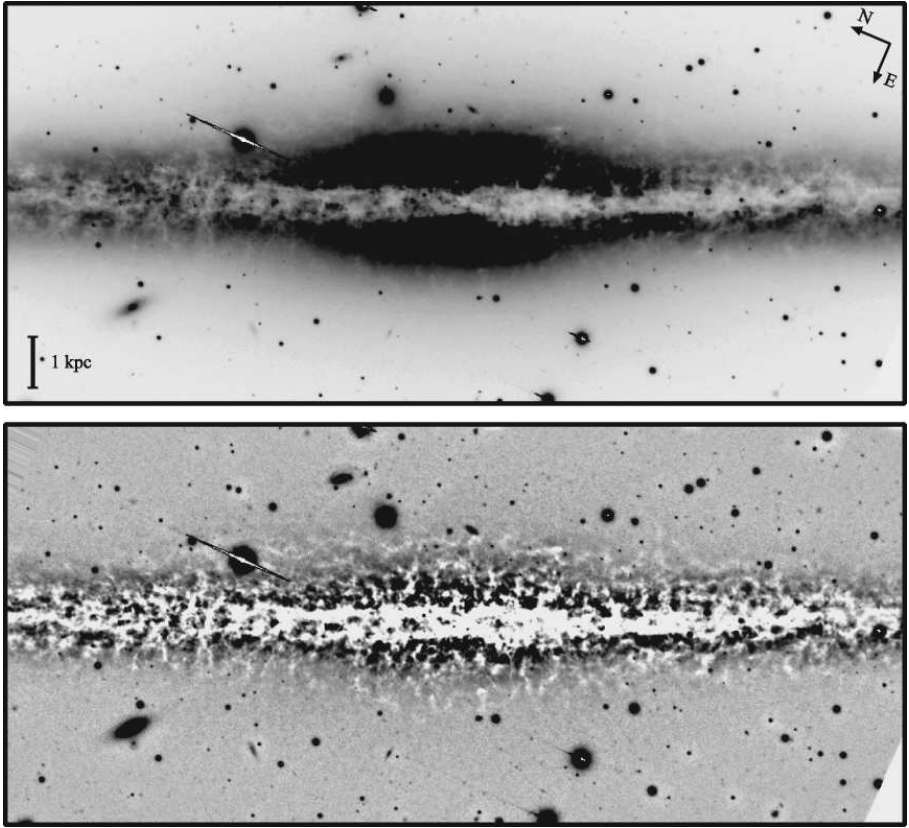}
\caption{Two versions of a broad-band optical (V-band) image of the
  edge-on galaxy NGC 891 from Howk \& Savage (2000).  The top panel
  shows the direct V-band image.  The bottom panel shows an
  unsharp-masked version of the V-band image.  The latter is produced
  by smoothing the original image and dividing the original by this
  smoothed version. This removes large-scale surface brightness
  variations, e.g., due to the vertically-decaying light from the
  stellar disk or from the bulge of the galaxy, and emphasizes small
  scale structures ($<300$ pc).  The clouds seen in this image each
  have masses in excess of $\sim10^5$ \msun; the total mass of the
  thick disk phase traced by this dust may be $\sim10^8$ \msun\ (Howk
  \& Savage 2000).  \label{fig:n891}}
\end{figure}

The evidence for extraplanar dust is not limited to the Milky Way.  We
have demonstrated in a series of papers (Howk \& Savage 1997, 1999a,
2000; Thompson et al. 2004; Howk 2005) that extraplanar dust can be
found in a large fraction of spiral galaxies (see also Rossa \&
Dettmar 2003, Alton et al. 2000, Keppel et al. 1991).  The evidence
for this comes from direct optical imaging of edge-on galaxies, as
shown in Figure \ref{fig:n891}, that reveal dust-bearing clouds the
background stellar light to heights approaching 2 kpc in many galaxies
(Howk \& Savage 1999a, Rossa \& Dettmar 2003).  

The clouds seen in images like Figure \ref{fig:n891} are only visible
because they are denser than their surroundings -- a smooth component
of dust will not be visible as small-scale absorption in these images.
An analysis of the properties of the dusty clouds (e.g., Howk \&
Savage 2000) suggests they each contain more than $\sim10^5$ \msun\ of
material and have densities in excess of 1 to 20 cm$^{-3}$ (Howk
2005), and there is evidence for on-going star formation in the thick
disks of several galaxies (Howk \& Savage 2000, Rueff et al. 2009).
This has led us to conclude that the dusty clouds seen using direct
optical imaging are likely to trace a cool, neutral medium in the
thick disks of spiral galaxies, perhaps one that harbors star
formation (Howk 2005; Howk \& Savage 2000).

This approach to detecting dust in the thick disk does not provide
much quantitative information on the distribution of grains in the
thick disk.  The detection of dusty thick disk clouds is strongly
biased towards clouds on the front side of a galaxy that are
significantly denser than their surroundings.  This method provides no
access to a diffuse distribution of grains.  The sight line confusion
due to the strong obscuration at $z<1$ kpc in most galaxies makes it
extremely difficult to connect the dust-bearing structures to activity
within the disk.  At high \z, it may become difficult to detect the
dust either because there is a lack of cool clouds which provide large
contrast (i.e., the dust is more diffuse), or the background stellar
light against which we search for the absorbing structures becomes too
faint.  Our analysis suggests the latter is not the case in some
galaxies, and we have no limits on the former through this technique.


\subsection{Diffuse Extraplanar PAH Emission}

With modern mid-/far-infrared instrumentation, it is now possible to
study thermal or fluorescent emission from dust grains in the thick
disks of nearby spiral galaxies.  Directly measuring the emission from
the grains in wavebands where the dust is optically thin allows us to
overcome some of the limitations of the direct optical imaging
discussed above.  

We have embarked on a program to study the 8$\mu$m band emission from
polycyclic aromatic hydrocarbons (PAHs; Tielens 2008) in the disk-halo
interface in a sample of edge-on spirals in the local Universe.  PAH
molecules represent the dominant heating source of the diffuse ISM, so
their presence is important to the thermal balance of the extraplanar
gas.  They are stochastically heated by UV photons, although they are
quite easy to excite and respond even to B stars.  Thus, we do not
expect the heating rate to be overly sensitive to the distance from
the plane within the first several kpc (e.g., Wolfire et al. 1995).

\begin{figure}[!h]
\includegraphics[width=4.75in]{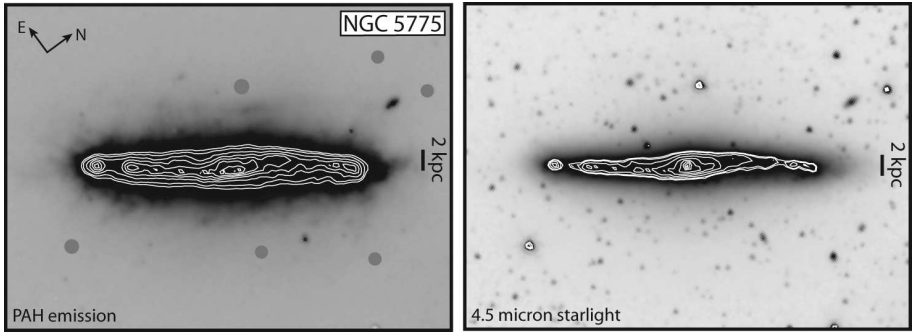}
\caption{\spitzer\ observations of the nearly edge-on galaxy NGC 5775
  showing PAH emission ({\em left}) in the 8 \micron\ band and the
  distribution of stellar light ({\em right}) in the 4.5 \micron\
  band.  A small stellar continuum contribution to the 8 \micron\
  image has been removed by scaling the 4.5 \micron\ image (see Helou
  et al. 2004) to show the pure PAH emission.  Filamentary PAH
  emission can be seen in this galaxy to $z\sim5$ kpc projected height
  from the midplane.  A diffuse component seems also to be present.
  We have found similar extraplanar PAH emission is seen in several
  other galaxies. Each image is $5 \arcmin \times 3.5\arcmin$ on the
  sky. \label{fig:n5775pah}}
\end{figure}

PAHs have been reported in the thick disks of normal spirals and
starbursts based on observations from both the {\em ISO} and \spitzer\
missions (Irwin \& Madden 2006, Engelbracht et al. 2006, Irwin et
al. 2007, Rand et al. 2008, Whaley et al. 2009).  We are expanding on
these efforts in a systematic way.  Using \spitzer 's IRAC intrument,
we have detected extraplanar PAH emission in several galaxies known to
harbor extraplanar dust and DIG.  Figure \ref{fig:n5775pah} shows the
\spitzer\ view of PAHs in the edge-on spiral NGC 5775.  One can trace
PAH-emitting filaments in this galaxy to $z\sim5$ kpc (projected).

The important conclusions so far from our work include: 1) Filamentary
PAH emission is seen to quite large distances, with galaxies showing
emission to $z\sim2$ to 5 kpc, depending on the galaxy; 2) A smooth
component of PAH emission seems to be present; 3) The scale heights of
thick disk PAH emission are of order $\sim0.7$ to 1.0 kpc in several
galaxies, similar to what is seen in the diffuse ionized gas (DIG); 4)
The PAH emission shows structures associated with previously seen
extraplanar dust {\em and} DIG structures.

The last point is important, as it not only implies that PAHs are
robust to the processes that lift this material out of the thin disk,
but they are also present and available for photoheating of the
neutral {\em and} ionized gas to large distances from the plane.  It
is not clear whether or not this would be able to provide significant
extra heating as suggested by some forbidden line observations of the
DIG (see discussion in Rand et al. 2008).  


\section{On the Mixture of Infalling Gas in the Thick Disks of Spiral
  Galaxies}
\label{sec:dust2gas}

Fraternali (2009, these proceedings) and collaborators (Fraternali \&
Binney 2008) have argued that the \HI\ rotation curves (Oosterloo et
al. 2007) for gas above the plane of NGC 891 are best explained by the
addition of low angular momentum {\em neutral} material into the
emitting regions.  While the majority of the \HI -emitting gas at
high-\z\ is likely still associated with galactic fountain-type
circulation, the 10\%\ or so of the gas that is newly accreted in
their model is flowing in at a rate nearly equal to the star formation
rate.

One potential way to test this and determine the mixture of ejected
versus accreted gas as a function of height is to study the ratio of
IR emission to \HI\ column density with \z.  As dust-rich material
from the thin disk mixes with dust-free accreted material, the dust to
gas ratio decreases relative to gas in the thin disk, as does the IR
flux/\HI\ column density ratio.  

While the mixing of primordial material into the interstellar medium
is the most exciting mechanism for changing the observed IR/\HI\ ratio
within the thick disk of a spiral galaxy, it is not the only process
that can affect this ratio.  We note the following complications to
applying this idea:

\begin{itemize}
\item {\bf The heating (radiation field) varies with height.} The
  local radiation field important for heating dust grains will
  diminish with height above the plane.  However, below $z\sim
  R_{disk}$ the radiation field does not vary too much with height
  once above the optically thick layer of dust associated with the
  thin disk (e.g., Wolfire et al. 1995).  Radial variations will also
  exist (see, e.g., the radiation fields presented in Fox et
  al. 2005), although much of the extraplanar dust is relatively
  confined in radius.  Thus, it seems likely a model of the UV
  radiation field required for heating could be constructed to account
  for the changes in the heating with position in the thick disk.  

\item {\bf The mixture of grains may change with height.}  Radiation
  pressure, dust destruction, and other processes may change the size
  distribution of grains with height, which can affect the
  distribution of temperatures and emissivity of the grains with \z.
  Wide wavelength coverage will be needed to probe the full range of
  grain sizes and temperatures.  For example, long wavelength IR
  observations will be needed to search for cool grains at high-\z,
  where large grains relatively far from heating sources may contain
  significant mass.  If a reasonable approximation to the radiation
  field can be made, multiwavelength observations may be able to shed
  some light on the grain size distributions and their changes with
  height.  

\item {\bf Dust may be injected in situ by AGB stars.}  If AGB stars
  contribute significantly to the mass of extraplanar dust, they can
  have a strong effect on the vertical distribution given their large
  scale height in galaxies.  Our early estimates suggest this will not
  be too large an issue.  The AGB contribution can be estimated with
  dust production models and direct observations of dust in AGB winds
  together with an observed distribution of AGB stars in the galaxies
  being studied.

\item {\bf Pristine material or low dust-to-gas ratio from dwarf
    galaxies may be accreted.}  Any gas with low dust-to-gas ratio
  compared with material ejected from the disk will lower the IR/\HI\
  ratio compared with the thin disk emission.  Continuity
  considerations suggest the fractional contribution from infalling
  matter to the thick disk gas should be highest at large \z.  

\end{itemize}

While some data from \spitzer\ exist that could in principle be used
to probe the IR/\HI\ ratio as a function of \z, a full accounting of
the IR emission will likely require future instruments, e.g., {\em
  Herschel} and {\em SOFIA}.  The limitations for \spitzer\ data
include wavelength coverage (especially accute given the limited
sensitivities at longer wavelengths) and background subtraction issues
in both IRAC and MIPS data that limit the ability to study low surface
brightness emission.  However, the lagging \HI\ rotation is seen in
NGC 891 at heights as low as $z \sim 4-5$ kpc.  These heights may be
accessible to \spitzer\ observations for some nearby galaxies.  

In addition, the nature of anomalous velocity clouds or high column
density clouds at large projected heights, such as the 22 kpc filament
seen in NGC 891 (Oosterloo et al. 2007), could be tested using the
IR/\HI\ ratio.  For example, one could test the origins of HVCs
projected onto the disks of spiral galaxies (see Sancisi et al. 2008):
they should have low IR/\HI\ ratios if they are falling onto a galaxy
from the IGM or represent matter stripped from a low-metallicity dwarf
galaxy.  If they have recently been ejected from the disk of the
galaxy, the ratio may not be significantly different than the average
of the  disk emission.

While there are significant difficulties to overcome in order to use
the IR/\HI\ ratio as a probe of the mixture of infalling primordial
gas with material ejected from the thin disks, the idea provides one
of the few ways to estimate this mixture as a function of position
within galaxies.  However, the identification and study of cool,
infalling matter in galaxies is extremely important given the
important role such accretion plays in the evolution of galaxies.

\vspace*{0.5cm} I thank the conference organizers for inviting me to
give this talk.  This work is funded by NASA via a research support
agreement through the Jet Propulsion Laboratory (RSA 1287864).  Thanks
also to my collaborators, M. Ashby, N. Lehner, K. Rueff, and B. Savage.


\end{document}